\documentclass[twocolumn,aps,prl,10pt,amsmath,amssymb,nofootinbib,showpacs,superscriptaddress,floatfix]{revtex4-1}

\DeclareFontFamily{U}{rcjhbltx}{}
\DeclareFontShape{U}{rcjhbltx}{m}{n}{<->rcjhbltx}{}
\DeclareSymbolFont{hebrewletters}{U}{rcjhbltx}{m}{n}

\usepackage{graphicx}
\usepackage{color}
\usepackage[usenames,dvipsnames]{xcolor}
\usepackage[colorlinks=true,linkcolor=Red,citecolor=Green,linktoc=page]{hyperref}
\usepackage{multirow}
\usepackage{float}
\usepackage{flushend}
\usepackage{balance}
\usepackage[varg]{txfonts}
\usepackage{ulem}
\usepackage{fancyhdr}

\DeclareMathSymbol{\lamed}{\mathord}{hebrewletters}{108}

\begin{document}
\title{Dark matter and dark energy in combinatorial quantum gravity}



\author{C.\,A.\,Trugenberger}

\affiliation{SwissScientific Technologies SA, rue du Rhone 59, CH-1204 Geneva, Switzerland}



\begin{abstract}
We point out that dark matter and dark energy arise naturally in a recently proposed model of combinatorial quantum gravity. Dark energy is due to the ground-state curvature at finite coupling, dark matter arises from allotropy in the discrete structure of space-time. The stable structure of the space-time ``crystal" represents the curved background, the coexisting metastable allotropes of higher curvature and energy are natural candidates for dark matter. We thus suggest that dark energy and dark matter are two manifestions of quantum gravity. 
\end{abstract}
\maketitle

\section{Introduction}

Dark matter (for a recent review see \cite{arbey2021}) and dark energy (for a recent review see \cite{brax2018}) are the two major puzzles of contemporary cosmology. The mass-energy of the universe is made only by 5\% of ordinary baryonic matter, the remainder being 27\% dark matter, an unknown form of matter interacting with ordinary matter only through gravity and 68\% dark energy, an unknown form of energy causing the measured acceleration of the universe expansion (for a review see \cite{peebles2003}). Even after decades of dedicated efforts, no proposed theory matches the data. 

The main candidates for the explanation of dark energy are a cosmological constant, a background negative pressure driving the universe expansion (see \cite{peebles2003, brax2018, padmanabhan2003}) or quintessence \cite{wetterich1988}, a dynamical version of the cosmological constant representing a fifth fundamental force. 

There are two possible avenues to tackle the dark matter problem. One is to posit a new type of particle, with essentially only gravitational interactions, which is the mainstream belief today, the other is to posit a modification of general relativity (GR). In this latter approach, mostly classical models have been proposed, like modified Newtonian dynamics (MOND) \cite{milgrom1983} or its relativistic generalization, tensor-vector-scalar gravity (TeVeS) \cite{bekenstein2004}. Apart from the dark matter and dark energy problems themselves, however, no other classical alteration of GR is required. What is called for, instead is a theory of quantum gravity. If dark matter is not matter, it is most likely that it is explained by the generalization of GR required to reconcile it with quantum mechanics. A first proposal in this sense is the so-called entropic gravity \cite{verlinde2011}, a model in which space time is considered as an emergent property of entangled qubits. 

It is the purpose of this paper to show that both dark energy and dark matter are natural consequences of quantum gravity. Based on recent advances in discrete geometry, we recently proposed to formulate general relativity on abstract metric spaces, focusing, specifically on random graphs \cite{tru2017, kelly2019, kelly2021}. The Hamiltonian governing this model is Ollivier's combinatorial Ricci curvature \cite{ollivier2007, ollivier2009, linn2011, jost2014}, hence the name ``combinatorial quantum gravity". Two of the characteristic features of this model are that there is no real distinction in the nature of space-time and matter, the two being just two different phases of the same elementary constituents, and that quantum mechanics in three spatial dimensions emerges simultaneously with gravity, in the spirit of Wheeler's ``it from bit" hypothesis \cite{tru2022, tru2023a, tru2023b}. As we now show, both dark energy and dark matter emerge naturally in this theory. 

\section{Hyperbolic holography}
The model posits a negative-curvature surface, which can be thought of as a curved holographic screen \cite{thooft, susskind1995, bousso2000}, and a universal Newtonian time parameter $t$ which governs its statistical fluctuations, including the Brownian motion of point-like defects. At distances larger than the curvature radius, Brownian motion on a negative-curvature surface decomposes into a dominant ballistic and deterministic component and a random correction \cite{tru2022, tru2023a, tru2023b}. The dominant deterministic component is equivalent (up to exponentially small corrections) to the geodesics of a de Sitter surface of positive curvature of the same absolute value, i.e. to the expansionary drag of a positive cosmological constant \cite{brax2018, peebles2003, padmanabhan2003} . The radial coordinate of the surface becomes dynamically soldered to universal time and emerges as coordinate time. 

When this emergent coordinate time is used to parametrize dynamics, the residual slow, stochastic component of Brownian motion, as seen by co-moving observers, is quantum mechanics in an emergent space with 3 spectral space dimensions and one time dimension. The spectral space dimension 3 is a consequence of the local limit theorem in negative curvature \cite{ledrappier} and goes by the name of ``pseudodimension" or ``dimension at infinity" in the mathematical literature \cite{anker2002}. The distances in this 3D emergent space are inherited from the hyperbolic surface and, as a consequence the effective positive curvature effects are reproduced also in 3D (see \cite{tru2023b} for a review). 

Overall, the picture of the universe in this model is as follows. At scales larger than the curvature radius of the holographic surface we have an effective (3+1)-dimensional de Sitter space with the slow, co-moving dynamics of massive particles governed by quantum mechanics. At scales smaller than the curvature radius, Lorentz invariance and the very concept of coordinate time are lost and the model reduces to diffusion on an effectively flat 2D surface. At even smaller scales, below the Planck scale, we see the discrete nature of this surface, as we will discuss in the next section. 

It is important to stress that, as a consequence of the hyperbolic holography, the curvature effects in this ``emergent universe" reduce to the fundamental curvature effects on the holographic surface. This will the subject of the rest of the paper.

\section{Emergence of 2D geometry from networks}

The idea of formulating a discrete version of general relativity goes back a long time. Most attempts are based on simplicial complexes, either causal ones \cite{loll2012} or growing flavoured ones \cite{bianconi2016, bianconi2017}. Simplicial complexes, however are still piece-wise flat chunks of Euclidean space. In \cite{tru2017, kelly2019} we proposed to formulate general relativity on abstract metric spaces, with no reference whatsoever to concepts such as a manifold, not even locally, focusing in particular on random graphs (for a review see \cite{albert2002}). To this end we used a then recently developed combinatorial form of Ricci curvature, the Ollivier curvature \cite{ollivier2007, ollivier2009, linn2011, jost2014}. A modified version of the Ollivier curvature was also proposed as a quantum version of Ricci curvature in \cite{loll2018a, loll2018b}. The Ollivier Ricci curvature, however, is the only combinatorial curvature (to our best knowledge) that has been proven to converge to continuum Ricci curvature for geometric networks defined on manifolds \cite{krioukov2021, kelly2022}. For another recent attempt to obtain an emergent space-time from purely random structures see \cite{kleftogiannis}. 

The holographic surface is formulated as a statistical mechanics model on the configuration space of 4-regular, incompressible random graphs (IRG) with $N$ vertices \cite{tru2017, kelly2019}. The partition function ${\cal Z}$ is
\begin{equation}
{\cal Z} = \sum_{IRG} {\rm e}^{-{1\over g} H} \ ,
\label{comqg}
\end{equation} 
where $g$ is the dimensionless coupling. Incompressible graphs are those for which short cycles, i.e. triangles, squares and pentagons, do not share more than one edge. This condition is the loop equivalent of the hard-core condition for bosonic point particles. As the hard-core condition prevents the infinite compressibility of Bose gases, the graph incompressibility condition prevents graph crumpling by requiring that loops can ``touch" but not ``overlap" on more than one edge. The graph incompressibility condition can be also formulated as an excluded sub-graph condition \cite{kelly2019}. The Hamiltonian $H$ is the total Ollivier-Ricci curvature, a discrete combinatorial analogue of the Riemannian Einstein-Hilbert action, 
\begin{equation}
H = -{4} \sum_{i \in G} \kappa (i) = -{4}\sum_{i\in G} \sum_{j\sim i} \kappa (ij) \ ,
\label{deh}
\end{equation}
where we denote by $j \sim i$ the neighbour vertices $j$ to vertex $i$ in graph $G$, i.e. those connected to $i$ by one edge $(ij)$. Here $\kappa (ij)$ is the Ollivier-Ricci curvature of edge $(ij)$ \cite{ollivier2007, ollivier2009, linn2011, jost2014} (see Appendix) and $4$ is the degree of the regular graph. 

While the Ollivier-Ricci curvature is cumbersome to compute in the generic case, it simplifies substantially on 
2D-regular incompressible graphs \cite{kelly2019}, 
\begin{equation}
\kappa(ij)= \frac{T_{ij}} {2D}-\left[1-\frac{2+T_{ij} +S_{ij}}{2D}\right]_+ - \left [1-\frac{2+T_{ij} +S_{ij} +P_{ij}}{2D}\right]_+ \ .
\label{orc}
\end{equation}
where the subscript ``+" is defined as $[\alpha]_+ = {\rm max} (0, \alpha)$ and $T_{ij}$, $S_{ij}$ and $P_{ij}$ denote the number of triangles, squares and pentagons supported on egde $(ij)$. This gives
\begin{eqnarray}
H &&= H_{\rm global} + H_{\rm local} \ ,
\nonumber \\
H_{\rm global} &&= 16 \left( N - {9\over 8} T - S- {5\over 8} P \right) \ ,
\nonumber \\
H_{\rm local} &&= \sum_{(ij) \in E_1} \left( (T_{ij} + S_{ij} ) - 2 \right) 
\nonumber \\
&&+ \sum_{(ij) \in E_2} \left( (T_{ij} + S_{ij} +P_{ij}) - 2 \right) \ ,
\label{globallocal}
\end{eqnarray}
where $T$, $S$ and $P$ are the total numbers of triangles, squares and pentagons on the graph and $E_1$ and $E_2$ are the ensembles of edges for which the respective summands are strictly positive. If we consider edges supporting each no more than two short cycles, the local term can be neglected. 

Let us consider, for simplicity, this case so that we can focus on the global term. This shows that the formation of short cycles is favoured since it minimizes the energy, with triangles being most favoured, followed by squares and lastly pentagons. The incompressibility condition, however, requires that a triangle can share an edge only with a pentagon, not with an another triangle or a square. But an edge supporting two squares contributes less to the energy than an edge supporting a triangle-pentagon couple. As a consequence, squares are favoured and triangles and pentagons can survive at most as rare isolated defects. Numerical simulations fully confirm this simple argument \cite{kelly2019}. From now on it will be simpler, thus to consider only bipartite graphs, which have no odd cycles at all.

Random regular graphs have essentially no short loops, these being exceedingly rare in large graphs \cite{wormald2004}. The ground state of the statistical model, instead is reached when the number of squares takes the maximal value $S=N$. One should thus expect a phase transition when $g$ is decreased from $g=\infty$, where the contribution of the highest-energy random regular graphs is large to $g=0$, where only the ground state survives. 

There is indeed a continuous phase transition between a random phase, with extremely sparse cycles on the graph and a geometric phase, in which the graph defines the tiling of a negative curvature surface (for a review see \cite{tru2023b}). The phase transition is due to a condensation of square loops (4-cycles), with triangles and pentagons appearing only as possible rare defects \cite{kelly2019}, as the dimensionless coupling constant (playing the role of temperature) is decreased. For a finite number of vertices, this is a genus-decreasing transition from an infinite genus surface at the critical point to a torus ground state at $g=0$. For an infinite number of vertices, we have a transition from infinite curvature at the critical point to a flat Euclidean plane at $g=0$. For all intermediate coupling we have a negative-curvature surface with two scales: an ultraviolet scale, below which the random character survives, and an infrared radius of curvature. When the coupling is reduced, both the ultraviolet scale and the curvature decrease, until a uniform flat plane is obtained as the ground state at zero coupling. 

The ultraviolet scale is the correlation length of the continuous phase transition. As always, below this scale, the disordered phase, in this case the random phase, survives.
In this graph transition, however, this has an additional consequence. In the geometric phase, distances scale as $\sqrt{N}$ while in the disordered phase, where a random graph survives, distances scale as ${\rm log}\ N$ \cite{albert2002, wormald2004}. When $N$ becomes very large, the dimensions of a disordered random ball, as viewed from the geometric phase, becomes point-like. Therefore, the surviving domains of random phase at any finite coupling have a natural interpretation as Planck-size matter particles, with a rest energy proportional to their excess curvature with respect to the ground state \cite{tru2022, tru2023a}. It is these point-like objects, which we shall call baryonic matter, that diffuse on the surface as a consequence of statistical fluctuations, as describe above. In this model, there is no real distinction between matter and space-time, both are made of the same constituents, graph links.

To conclude this section, it is interesting to note that most discrete models of gravity based on networks produce negative-curvature ground states. Indeed, this result was found both in growing simplicial complexes approaches \cite{bianconi2017} and in models based on an alternative combinatorial Ricci curvature \cite{loll2018a}, although the authors of this latter work chose to interpret this as ``quantum flatness".

\section{Dark energy} 
Dark energy, in the present model, is a simple consequence of ground state curvature at finite coupling. In the geometric phase, each vertex $i$ can be surrounded by $s_i=1,2,3,4$ squares (we have already treated the $s_i=0$ case in the previous section). The minimum of the free energy at a given coupling $g$ is realized for one of these values, which we shall call $\bar s$ and which is the most frequent value for this $g$. Let us consider a uniform configuration with $\bar s$ at each vertex and all edges of fixed length $\ell$. As we discuss in detail in the next section, this is the tiling of a hyperbolic plane with a given curvature dependent on $\bar s$. This curvature, by means of the hyperbolic holography described above, plays the role of a cosmological constant \cite{padmanabhan2003}, i.e. of dark energy. We will discuss the role of deviations from the most frequent value $\bar s$ in the next section. 

If we consider the coupling $g$ as a dynamical variable, it is possible to identify $1/g$ itself with universal time.  In this case, the big bang can be identified with the critical point at which the universe emerges as an infinite-curvature point-like ball of baryonic matter. From there on, the curvature decreases from extremely high values towards zero while more and more matter is transformed into space-time. Today, we expect the radius of curvature scale to lie between the Planck scale and the smallest scales probed by present accelerators: below this scale one would see 2D physics. The very large curvature near the critical point would be perceived at large distances in the effective de Sitter description as an extremely accelerated expansion, a form of topological inflation when space-time emerges from randomness by a topological re-arrangement of the graph. Of course, one should also take into account the gravitational attraction between the residual matter. The details of such a computation are beyond the scope of the present paper but it is clear that this would slow considerably the perceived expansion acceleration, perhaps even cancelling it completely for a period. But when more and more matter transforms into space-time, so that the residual attraction becomes small, the ground-state curvature takes over again and the perceived expansion will accelerate again, albeit at a much smaller rate, which actually tends to zero as the ground state curvature becomes smaller and smaller at tiny couplings. The ground state curvature is thus a natural explanation of dark energy. 

\section{Dark matter}
Every vertex $i$ in the geometric phase can exist in four possible states, corresponding to the number $s_i=1,2,3,4$ of squares surrounding it, which, as we now show, amounts to a quantized curvature. We have identified network regions with $s_i=0$ with baryonic matter particles, while the value $\bar s$ realized in the minimum of the free energy at a given coupling defines what we call space and has the lowest curvature, giving rise to dark energy as explained above. But what about vertices with different values of $s_i$? 

As always when slowly lowering the coupling in continuous phase transitions, there typically survive domains of higher free energy. These are metastable states, which, depending on their free energy difference to the minimum, may be extremely long-lived. In the present case, these domains interact gravitationally due to their higher curvature. Such domains, characterized by $0 <s_i <\bar s$ are neither baryonic matter nor space but, rather, represent natural candidates for dark matter.  Dark matter, in this model, appears like crystal allotropy (for a review see \cite{bernstein2002}) in the fabric of space-time. 

Let us now explain this in details. Every regular graph with degree larger than 2 has a genus, corresponding to the smallest genus of a surface on which it can be embedded with no edge crossings. The graph becomes thus the 1-skeleton of a 2-cell embedding, for which the graph cycles are homeomorphic to open disks on the surface. The 2-cell embedding is called a ``map”. A combinatorial map can be geometrized by assigning the same fixed geodesic length $\ell$ to each edge so that it becomes a tiling of the surface, in which the combinatorial cycles become regular polygonal faces of the tiling \cite{datta2019, maiti2020, maiti2023}. Here we shall consider infinite graphs in the geometric phase and, for simplicity, we shall focus on the geometric structure at large distances, approximating the baryonic matter domains with points. 

Sufficiently deep in the geometric phase, each edge of the map supports exactly two regular polygons. Let $k_i$, $i=1\dots 4$, be the number of edges of the 4 polygons surrounding a vertex and let us call the vector ${\bf k} = (k_1, k_2, k_3, k_4)$ the vertex type. For simplicity, we shall consider homogeneous maps, for which the edge type is the same for all vertices, so that the geometrization gives rise to semi-regular tilings.
Moreover, for homogeneous maps, the tiled surface is a constant curvature surface \cite{datta2019, maiti2020, maiti2023}. 

The combinatorial Ollivier-Ricci curvature of such a homogeneous map is given by (see (\ref{globallocal}) 
\begin{equation} 
\kappa_{\rm comb} = -16 N \left( 1-{9\over 8} {T\over N} -{S\over N} -{5\over 8} {P\over N} \right) = -H  \ ,
\label{combcurv}
\end{equation}
where $T$, $S$ and $P$ are the total numbers of triangles, squares and pentagons on the graph, and $H$ is its energy.  This has to be compared with the geometric curvature of the corresponding semi-regular tiling, given by the angle sum parameter 
\begin{equation}
\alpha = \sum_{i=1}^4 {k_i-2\over k_i} \ .
\label{angsum}
\end{equation}
If $\alpha <2$ the surface is spherical, i.e. of positive constant curvature, if $\alpha=2$ it is the flat Euclidean plane and, finally, if $\alpha >2$ it is hyperbolic, i.e. of negative constant curvature \cite{datta2019, maiti2020, maiti2023}. 

For every hyperbolic tiling there is exactly one geodesic length so that the sum of interior angles at each vertex sums exactly to $2\pi$. To see this, let us use the cosine rule for a hyperbolic n-gon,
\begin{equation}
{\rm sin} \left( {\theta \over 2} \right) = {{\rm cos} \left( {\pi \over n} \right) \over {\rm cosh} \left( {\ell \over 2R} \right)} \ ,
\label{hypercos}
\end{equation}
to write the total interior angle at a vertex of a tiling as
\begin{equation} 
\sum_{i=1}^4 2 \ {\rm arcsin} \left( {{\rm cos} \left( {\pi \over k_i } \right)  \over {\rm cosh} \left( {\ell \over 2R} \right)} \right) = 2\pi \ ,
\label{intan}
\end{equation}
where $R$ is the radius of curvature of the Poincar\'e disk on which the tiling is constructed. First of all let us note that this is a monotonically decreasing function of $\ell / R$. Second, for $\ell / R \to \infty$ the total interior angle vanishes. In the opposite limit $\ell / R \to 0$, the geometry becomes Euclidean and the interior angle sum becomes larger than $2\pi$ because of the above angle-sum condition $\alpha > 2$. Therefore, given a specific semi-regular tiling, there is exactly one possible parameter $\ell/R$ that solves (\ref{intan}). If the length unit $\ell $ is held fixed, the hyperbolic radius $R$ of the Poincar\'e disk varies and, with it, the curvature $K=-1/R^2$. Otherwise, we can describe varying curvature at a fixed radius by letting the length unit $\ell$ change. In any case, the tiling determines the curvature of the underlying surface. This shows also that, for strictly negative Ollivier curvature, we cannot take the length scale to zero; this is possible only if the curvature also vanishes. In the infrared, the model is pushed to the completely ordered phase at zero coupling, corresponding to a flat geometry. 

The geometric phase is realized by the condensation of squares and corresponds, thus to vertex types with at least one $k_i=4$ and all other $k_i= 4$ or $k_i \ge 6$ for all vertices. At each coupling, one of the four possible vertex types of the semi-regular tiling, with one, two, three of four squares will constitute the minimum of the free energy. When the coupling is decreased from the critical value, the number of squares corresponding to the free-energy minimum will increase, see (\ref{combcurv}). At each step, one of the $k_i$ decreases from a higher value to 4. In order to maintain the total interior angle sum as $2\pi$, as in (\ref{intan}), the quantity $\ell /R$ has to decrease, which means the absolute value of the curvature of the tiled surface decreases, until it vanishes for four squares. 

\begin{figure}[t!]
\includegraphics[width=7cm]{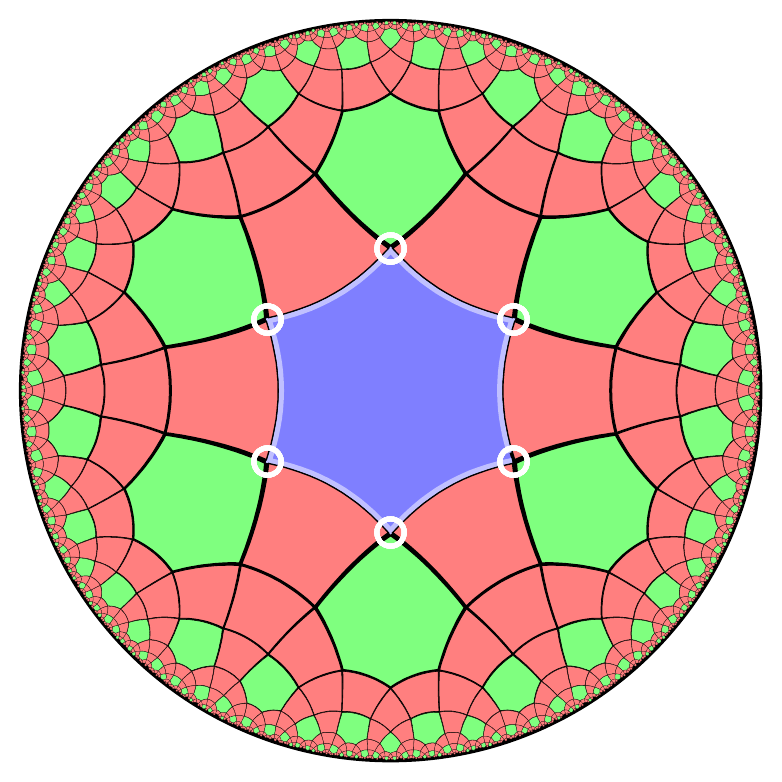}
\vspace{-0.3cm}
\caption{An allotropy region (blue) with vertices surrounded by two squares within an hyperbolic tiling with all vertices surrounded by three squares. The hyperbolic tiling represents space of a given curvature; the blue allotropy region represents a region of dark matter, with a higher absolute curvature and energy. Created by Eryk Kopczyński using RogueViz \cite{marek2017}.}
\label{Fig1}
\end{figure}

Suppose now the coupling decreases from the critical value to a lower value. If, at this new value, the free-energy difference between the actual minimum and the previous minimum at higher couplings is sufficently small, very long-lived metastable domains of a different tiling can survive, exactly as in crystal allotropy (for a review see \cite{bernstein2002}), the paramount example being unstable diamond in an environment which favours graphite as the stable configuration of carbon. These domains have higher absolute negative curvature than the stable ground state, as shown in the example in Fig.\,\ref{Fig1}. In the effective 3D Lorentzian picture arising from holography they have, correspondingly higher positive curvature and are thus natural candidates for dark matter. 

In this model, dark matter arises thus essentially as ``crystallographic" defects in the fabric of space-time. Of course, for couplings just above the transition values between two allotropes, small domains of lower absolute curvature can form. In the Lorentzian picture, these would correspond to antigravity domains with positive curvature lower than the dominant de Sitter background.

\section{Appendix: Ollivier-Ricci curvature}

Ricci curvature on manifolds is a measure of how much (infinitesimal) spheres around a point contract (positive Ricci curvature) or expand (negative Ricci curvature) when they are transported along a geodesic with a given tangent vector at the point under consideration. The Ollivier curvature is a discrete version of the same measure. For two vertices $i$ and $j$ it compares the Wasserstein (or earth-mover) distance $W\left( \mu_i, \mu_j \right)$ between the two uniform probability measures $\mu_{i,j}$ on the spheres around $i$ and $j$ to the distance $d(i,j)$ on the graph and is defined as
\begin{equation}
\kappa (i,j)= 1- {W\left( \mu_i, \mu_j \right) \over d(i,j)} \ .
\label{olli}
\end{equation}
The Wasserstein distance between two probability measures $\mu_i$ and $\mu_j$ on the graph is defined as
\begin{equation}
W\left( \mu_i, \mu_j \right) = {\rm inf} \sum_{i,j} \xi(i,j)d(i,j) \ ,
\label{wasser}
\end{equation}
where the infimum has to be taken over all couplings (or transference plans) $\xi(i,j)$ i.e. over all plans on how to transport a unit mass distributed according to $\mu_i$ around $i$ to the same mass distributed according to $\mu_j$ around $j$. 


\section*{Data availability}
Data sharing not applicable to this article as no datasets were generated or analyzed during the current study.

\section*{Competing interests}
The author declares that he has no competing interests.
	

\hskip 1pt









\end{document}